# Reduced effective scanning time in SPECT due to OSEM accelerated reconstruction

Krzysztof Kacperski [a), b)], Dominika Świtlik [a), c)], and Jakub Pietrzak [a), c)]

*Abstract–* We study the increase of image noise which can be observed in images reconstructed with the OSEM algorithm. We argue that the excessive noise is equivalent to shortening scanning time by a factor of 2 or more at high number of subsets (acceleration factors). Therefore, by keeping the number of OSEM subsets low one can significantly reduce image noise, which in turn is equivalent to increasing scanning time or radiation dose. While the OSEM remains a very useful reconstruction method enabling substantial reduction of reconstruction times, it should be used with caution. In particular, high numbers of subsets should be avoided if possible. Whenever the available computing power allows achieving acceptable reconstruction times at lower number of subsets, one can consider reducing scanning time or injected dose, while maintaining the image quality of standard scans reconstructed with higher OSEM acceleration. The recommended maximal "safe" number of subsets, which does not lead to significant increase of image noise, is about 5.

## I. Introduction

Ordered Subsets Expectation Maximisation (OSEM) algorithm has been a major development in statistical reconstruction methods [1]. The previously known Maximum Likelihood Expectation Maximisation (MLEM) algorithm has had prohibitively long computation times, which made it impractical for clinical applications. The OSEM uses a simple numerical trick in which the current image estimation is updated using a small subset of measured projections in a single iteration. The result of such iteration step is nearly equivalent to the update which uses all the measured projections, as in ML-EM [3]. Since the numerical computation of projections from current image estimate takes most of the computing time, acceleration factor which can be achieved in this way is nearly as high as the number of subsets, i.e. of the order of 10 - 20 or more. This was sufficient to bring the reconstruction times of iterative algorithms to clinically acceptable levels.

It has also been well recognised, that the OSEM algorithm needs to be used with caution. In fact, it does not guarantee convergence to a unique solution, and comes at a price of increased image noise and possible artefacts [1],[3]. Nevertheless, in most practical cases it works well, and has become a method of choice in numerous clinical scanners.

In this work we investigate noise properties of SPECT images reconstructed with OSEM algorithm with variable number of subsets and compare them to the original ML-EM algorithm. The image noise levels increased due to OSEM are matched by images acquired at reduced scanning times and reconstructed with ML-EM.

## II. Materials and methods

### A. Phantom

We used cylindrical numerical phantom of the diameter 42 cm filled uniformly with water solution of 200 MBq $^{99m}$Tc to simulate the background. 12 spherical hot lesions of diameters 17, 26, 33 and 50 mm and contrast ratios: 2:1, 3:1 and 4:1 have been placed symmetrically at the distance 12.5 cm from the central axis of the phantom in three layers.

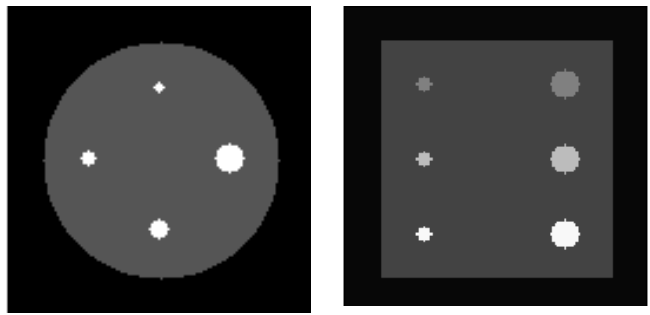

Fig. 1. Numerical Phantom with hot lesions: transaxial (left) and lateral (right) cross-sections.

One phantom contained small lesions of diameters 10, 14, 16 and 17 mm and tumour to background contrast ratios 5:1, 10:1 and 15:1 for each lesion size. The other phantom contained big lesions of diameters 17, 26, 33 and 50 mm of lower contrast ratios: 2:1, 3:1 and 4:1 respectively at the same positions.

### B. Scanning and reconstruction

SPECT scans with 120 projection angles over 360° using the low energy high resolution (LEHR) collimator have been simulated and the images reconstructed using the OSEM algorithm with variable number of subsets, as well as the plain ML-EM for comparison. Photon attenuation and collimator/detector point response function have been modelled both in projection and the reconstruction; Compton scatter has been neglected. Inter-iteration Gaussian smoothing has been used to control noise in the reconstruction. For each noise-free projection 30 instances of Poisson noise were generated to assess image noise. The count levels in projections were scaled to simulate different scanning times for a single detector head.

This work was supported by the Wellcome Trust under Grant No. 084288/Z/07/A.
a) The Maria Skłodowska-Curie Memorial Cancer Centre and Institute of Oncology, Warsaw, Poland.
b) National Centre for Nuclear Research, Świerk, Poland
c) Faculty of Physics, University of Warsaw, Poland.

email: Krzysztof.Kacperski@ncbj.gov.pl

The results below are shown for smoothing kernel with $\sigma = 2.5$ mm.

### C. Image quality measures

To quantify image noise 50 background masks of the same sizes as the lesions were created and placed randomly in 5 layers, 10 masks in each: three layers with the lesions and additional two symmetrically between them.

The following conventional parameters were computed for each set of scan and reconstruction parameters (cf. [8]):

Contrast:
$$C_l = \left\langle \frac{M_l - \overline{M_{bgd}}}{\overline{M_{bgd}}} \right\rangle$$

where $M_l$ is the average number of counts in the lesion, and $M_{bgd}$ – the average number of counts in the background masks. $\overline{M}$ denotes the average over the background masks within the layer of the lesion of a single image, and $\langle M \rangle$ is the ensemble average over the 30 noisy image instances.

Contrast recovery coefficient (CRC):
$$CRC = \frac{C_{rec}}{C_{org}}$$

average Coefficient of Variation (COV) in the background region:
$$COV = \frac{\overline{\sqrt{var\langle M_{bgd}\rangle)}}}{\langle M_{bgd}\rangle}$$

where $\bar{x}$ is the average over all the background masks.

### III. RESULTS

We use Gaussian smoothing of the image estimate before each next iteration to control noise. Smoothing kernel with $\sigma = 2.5$ mm yields close to optimal performance in terms of the maximal CNR for most lesions.

Figs. 1 and 2 show a typical behaviour of the contrast and noise parameters in the OSEM reconstruction. While the contrast essentially does not depend on the number of subsets, we can observe significant increase of image noise. This dependence is plotted in Fig. 3 for a range of scanning times. Note that the increase of noise due to the OSEM algorithmic acceleration can be matched by effective reduction of scanning time. For example the noise level in images obtained from 60 min. scans reconstructed with 10 subsets is approximately equal to that of 45 min. images reconstructed with ML-EM. Increasing the acceleration factor to 30 makes the noise exceed that of a 30 min. scan reconstructed with ML-EM. The effect is even stronger for shorter scanning times. Noise increase remains relatively low if the number of subsets does not exceed 5.

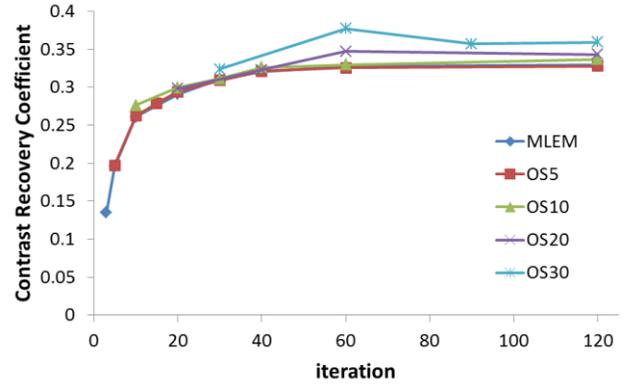

Fig. 1 CRC as a function of OSEM iterations for the 26mm 1:2 lesion; 60 min. scan.

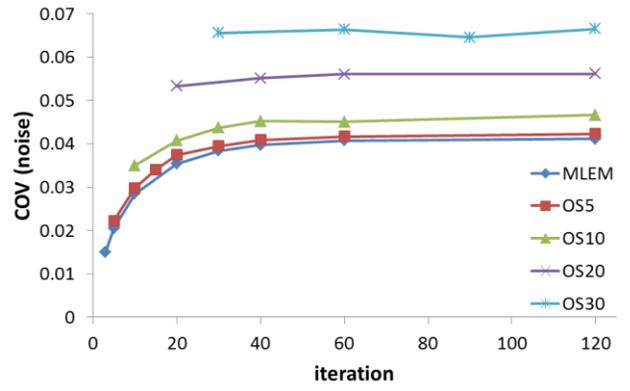

Fig. 2 COV as a function of OSEM iterations for the 26mm 1:2 lesion; 60 min. scan.

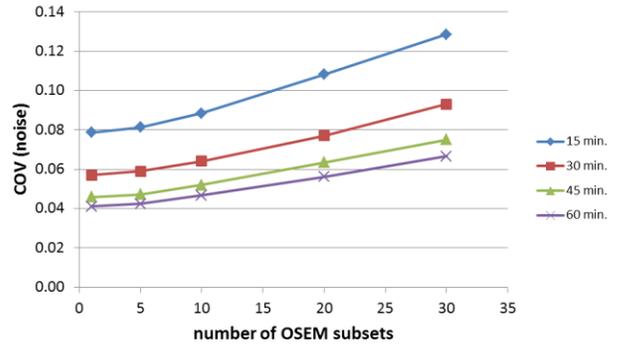

Fig. 3 COV as a function of OSEM iterations for different scanning times.

Contrast-noise curves (Fig. 4) confirm, that using 10 subsets in the OSEM reconstruction compromises image quality equivalently to reducing scanning time by 50% while using the ML-EM algorithm. Increasing the number of subsets to 20 corresponds to ML-EM image obtained at half the scanning time.

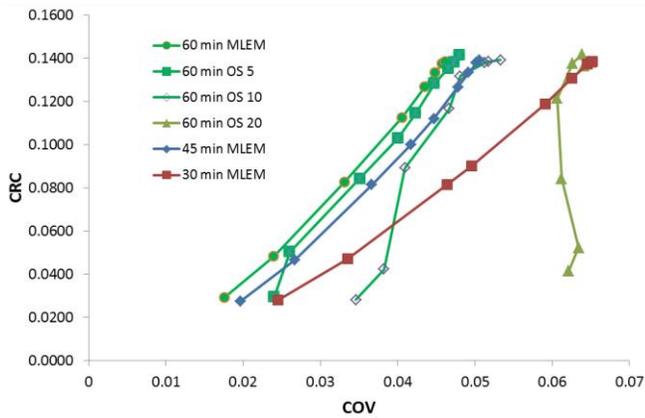

Fig. 4 Examples of contrast-noise curves for the 15 mm 1:2 lesion.

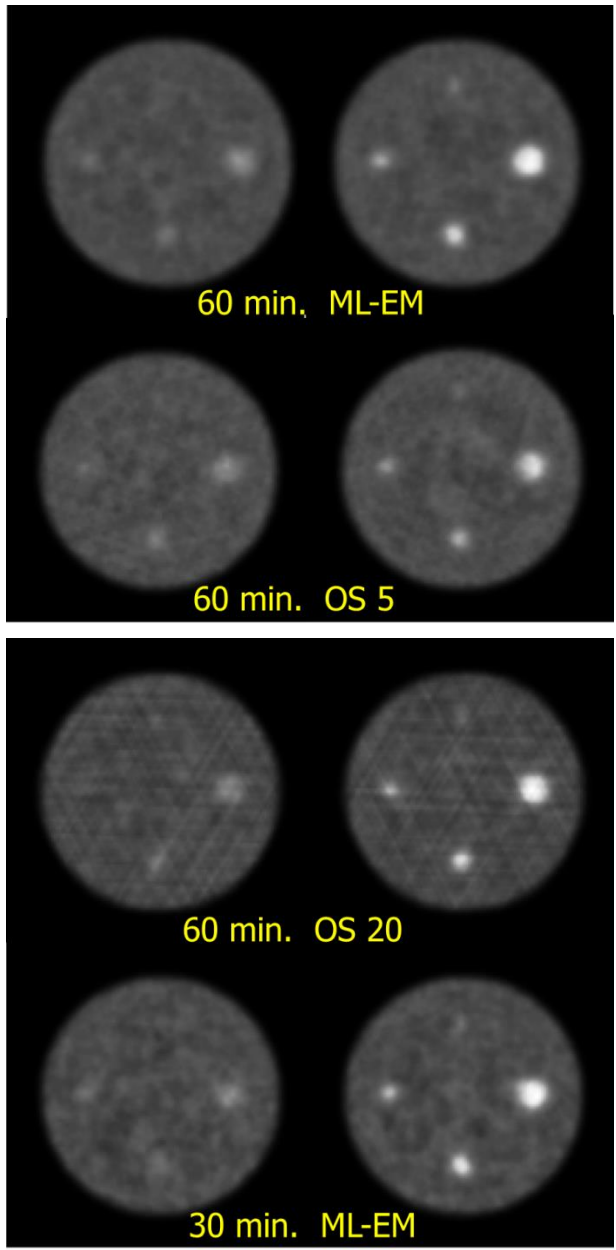

Fig. 5 Cross sections of the reconstructed phantom images showing the lesions of contrast 1:2 (left) and 1:4 (right)..

## IV. CONCLUSIONS

The OSEM algorithm offers a substantial acceleration of the maximum likelihood iterative reconstruction, however, at a price of increased image noise. This may be equivalent to reducing scanning time by 50 % while using the ML-EM reconstruction.

As the capabilities of modern computers, particularly extended by massive parallel processors, like GPU, have increased substantially over the recent years, the use of the OSEM algorithm, especially with high number of subsets, becomes questionable. When the number of subsets exceeds approximately 5 the increase of image noise becomes significant. Modest algorithmic acceleration can now be easily compensated for by e.g. massive parallel hardware implementation and the reduced image noise traded off for shorter scanning time or lower radiation dose. These are much desired to improve patient throughput and comfort as well as radiation safety.